\documentclass[conference]{IEEEtran}
\IEEEoverridecommandlockouts

\def\BibTeX{{\rm B\kern-.05em{\sc i\kern-.025em b}\kern-.08em
T\kern-.1667em\lower.7ex\hbox{E}\kern-.125emX}}
\usepackage[T1]{fontenc}      
\usepackage{hyperref}         
\usepackage{url}              
\usepackage{booktabs}         
\usepackage{amsmath}          
\usepackage{amssymb}          
\usepackage{mathtools}        
\usepackage{nicefrac}         
\usepackage{bm}               
\usepackage{microtype}        
\usepackage{xspace}           
\usepackage{enumitem}         
\usepackage{xcolor}           
\usepackage{cleveref}         
\usepackage{pifont}           
\usepackage{siunitx}          
\usepackage[symbol*]{footmisc} 

\usepackage{svg}
\usepackage{url}
\usepackage{cite}
\usepackage{algorithmic}
\usepackage{graphicx}
\usepackage{textcomp}

\renewcommand{\vec}[1]{\bm{#1}}
\newcommand{\mat}[1]{\bm{\mathrm{#1}}}
\newcommand{\widebar}[1]{\mkern 1.5mu\overline{\mkern-1.5mu\raisebox{0pt}[1.1\height]{$#1$}\mkern-1.5mu}\mkern 1.5mu}

\newcommand{\cmark}{\ding{51}} 
\newcommand{\xmark}{\ding{55}} 

\newcommand{\bigO}[1]{\mathop{\kern0pt \mathcal{O}}\left({#1}\right)}

\begin{document}

\title{SepMamba: State-space models for speaker separation using Mamba\\
}

\author{\IEEEauthorblockN{
Thor Højhus Avenstrup\IEEEauthorrefmark{1},
Boldizsár Elek\IEEEauthorrefmark{1},
István László Mádi\IEEEauthorrefmark{1},
András Bence Schin\IEEEauthorrefmark{1}, \\
{Morten M\o rup}\IEEEauthorrefmark{1},
{Bj\o rn Sand Jensen}\IEEEauthorrefmark{1} and
    {Kenny Olsen}\IEEEauthorrefmark{1}\IEEEauthorrefmark{2}}
\IEEEauthorblockA{\IEEEauthorrefmark{1}\textit{Technical University of Denmark}}
\IEEEauthorblockA{\IEEEauthorrefmark{2}\textit{WS Audiology A/S}}
}

\maketitle
\begin{abstract}
Deep learning-based single-channel speaker separation has improved significantly in
recent years in large part due to the introduction of the transformer-based attention
mechanism. However, these improvements come with intense computational demands,
precluding their use in many practical applications.
As a computationally efficient alternative with similar modeling capabilities, Mamba
was recently introduced. We propose SepMamba, a U-Net-based architecture composed of
bidirectional Mamba layers.
We find that our approach outperforms similarly-sized prominent models --- including
transformer-based models --- on the WSJ0 2-speaker dataset while enjoying significant
computational benefits in terms of multiply-accumulates, peak memory usage, and
wall-clock time. We additionally report strong results for causal variants of SepMamba.
Our approach provides a computationally favorable alternative to transformer-based
architectures for deep speech separation.
\end{abstract}


\section{Introduction}\label{sec:intro}

Speech separation is the problem of extracting separate source signals from a single
mixture, also known as the cocktail party problem~\cite{cherry1953experiments}, and is
a fundamental task in many modern audio processing systems, such as hearing aids or
telecommunication devices --- systems which are typically low-resource environments
making computationally expensive methods impractical.


Recent deep learning models have greatly improved the quality of speaker separation
systems, but at the cost of significant computational overhead. Prior approaches to
deep learning-based speaker separation have until recently chiefly been based on using
the short-time Fourier transform (STFT) to map input audio mixture into the
frequency-domain, performing separation on the (complex) frequency representation, and
then using the inverse STFT to construct the estimated
sources~\cite{isik2016singlechannel}. These methods come with several drawbacks.
Firstly, the STFT transforms the signal into the complex domain, where both magnitude
and phase must be explicitly modelled. In practice, using the phase information is
challenging, and without careful design models often end up reusing the input phase for
the estimated sources, which sets an upper bound on the separation
quality~\cite{tasnet}. Secondly, for high-quality speech separation large frame lengths
are typically used which heavily limits the usefulness of such models in cases where
low latency is required.

Instead of modeling the mixture signal in the time-frequency representation, several
recent approaches attempt to model the mixture directly in the time-domain. The first
time-domain solution to demonstrate superior performance to STFT-based systems was the
TasNet~\cite{tasnet}. TasNet proposed an encoder-masker-decoder architecture where
separation is performed in a learned linearly encoded basis, and decoded using another
linear basis, reminiscent of STFT-based systems but using real-valued learnable bases.
The masking network in the original TasNet was a deep long short-term memory (LSTM)
network, which was later replaced in Conv-TasNet~\cite{convtasnet} using only
convolutional layers throughout the whole model. Notably, Conv-TasNet outperformed all
existing solutions (in terms of SI-SNR and SDR) on the WSJ0-2mix dataset in the
non-causal setting and reached comparable results on the causal setting with a
substantially smaller model size. Later SudoRM-RF~\cite{tzinisSudoRmRf2020}, a
convolution-based model, proposed a resource efficient architecture with improved
performance over its predecessors.

\begin{figure}[t]
    \centering
    \includegraphics[width=\linewidth]{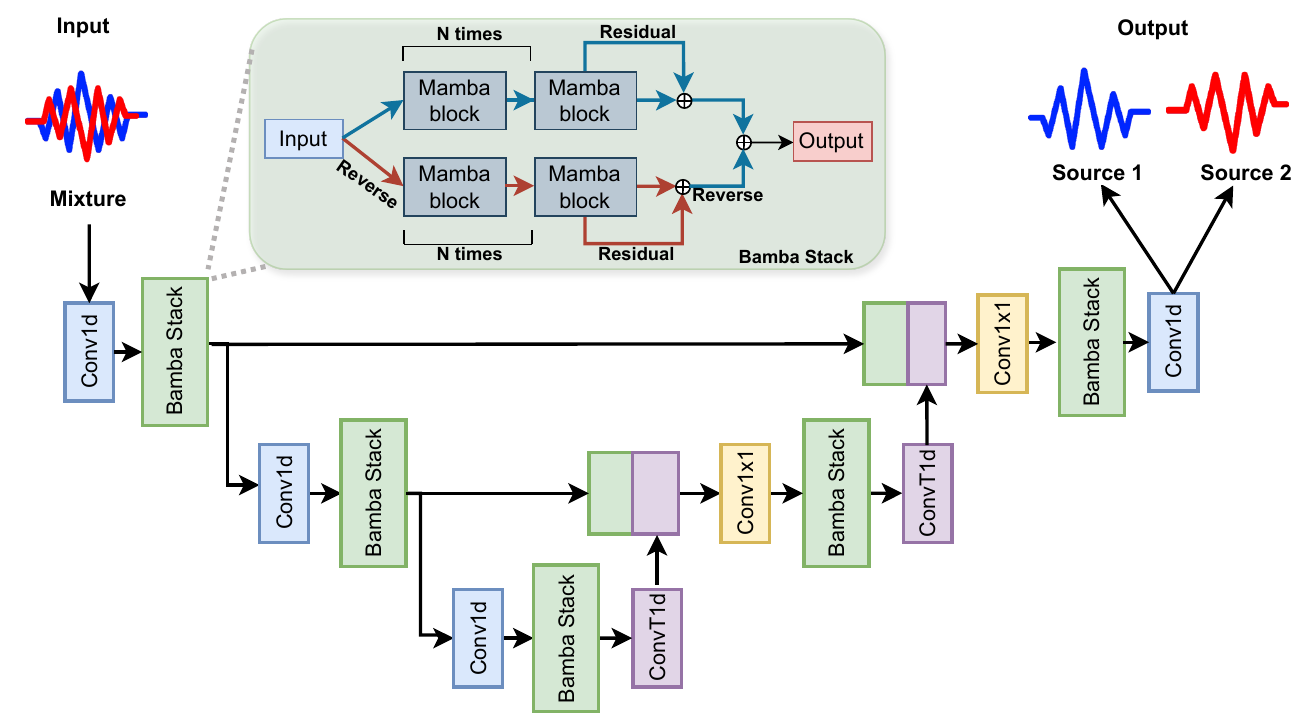}
    \caption{
        SepMamba has 5 stages of processing with Bamba stacks. The downsampling and
        upsampling is handled by convolutional and matching transposed convolutional
        layers. The skip connections are projected into the required dimension with $1
        \times 1$ convolutions. We double the dimension of the Mamba blocks after each
        downsampling by a factor of 2, and halve it after each upsampling (ensuring
        matching dimensions on the same level). }\label{fig:unet_structure}
\end{figure}

Transformer-based models have become the new standard backbone in deep speech
separation models~\cite{sepformer, mossformer, tfgridnet} due to their strong modelling
capacity and efficient use of parallelism on GPUs during training.
SepFormer~\cite{sepformer} was the first model to rely entirely on a transformer-based
masker network in a TasNet-like structure and achieved state-of-the-art performance on
the WSJ0-2/3mix dataset. Here, the masker network employs transformer blocks to capture
both short-term and long-term dependencies. Later MossFormer~\cite{mossformer} and
MossFormer2\cite{mossformer2} employed a unified attentive gating model, further
improving performance on the WSJ0-2/3mix datasets.

A notable drawback of these models, and generally of models building on the
transformer-based attention mechanism~\cite{vaswani2023attention}, is their quadratic
complexity over the input sequence length, which necessitates processing samples in
shorter chunks (e.g.\ SepFormer uses a chunk size of 250 samples). While this approach
effectively handles dependencies within and across chunks due to the dual-path
structure --- first introduced by DP-RNN~\cite{luoDualpathRNNEfficient2020} ---, it may
still struggle with long-range dependencies if critical information spans multiple
chunks. Furthermore, the process of segmenting the input during training, which is
counter-intuitive for audio --- an inherently continuous signal --- leaves the question
of whether sequence models that can model a full signal without segmentation could
further improve performance and efficiency.





State-space models have emerged as a promising alternative to existing sequence
modeling architectures. Recently, Mamba~\cite{mamba} has shown strong results in
language modeling and the modeling of DNA within genomics where it beat or matched
transformer-based architectures. Furthermore, Mamba has been explored in biomedical
image segmentation tasks~\cite{ma2024umamba} as well, where it also outperformed CNN-
and transformer-based segmentation networks. Notably, Mamba was also used for the
unconditional generation of audio~\cite{mamba}, using a Sashimi
architecture~\cite{goel2022its}, and was found to outperform its predecessor, the
S4~\cite{guEfficientlyModelingLong2021} layer.

Mamba has previously been combined with U-Nets, e.g., in the imaging
domain~\cite{wang2024mambaunet, liao2024lightmunet} and U-Net inspired hierarchical
structures with Mamba has also been successfully developed in the context of sequence
modeling for sensor based data~\cite{bhirangiHierarchicalStateSpace2024}. In the speech
separation space, SP-Mamba~\cite{spmamba} builds on the successful
TF-GridNet~\cite{tfgridnet} architecture by exchanging the bidirectional LSTM
components with bidirectional Mamba blocks. While SP-Mamba reports strong results, it
still employs the transformer-based attention mechanism, making it computationally
resource-demanding.

The rise of state-space models --- especially Mamba --- as a general sequential model
poses an opportunity to develop new, less resource-intensive and more computationally
efficient solutions in the speech separation space as well, while achieving performance
comparable to current state-of-the-art transformer-based models~\cite{sepformer,
mossformer, tfgridnet}.


We propose SepMamba, the first Mamba-based architecture for speaker separation in the
time domain that does not rely on expensive transformer-based attention mechanisms. The
new architecture achieves competitive performance in both causal and non-causal
settings by incorporating Mamba layers into a U-Net architecture to efficiently learn
multi-scale structures in sound, enabling inexpensive learning of long-range
dependencies. We provide a comprehensive overview of existing speaker separation
results on WSJ0-2mix and report strong performance, matching or outperforming
transformer-based architectures at a fraction of the computational costs. Additionally,
SepMamba achieves much lower forward and backward pass wall-clock timings and a
significantly lower peak memory use than comparable models.\footnote{The implementation
of SepMamba is available at: \url{https://github.com/andrasschin/SepMamba}.}

\section{Methods}


The Mamba layer is a sequence-to-sequence transformation mapping a 1-D input signal
with discrete-time samples $x_t\in\mathbb{R}$ to a 1-D output signal $y_t\in\mathbb{R}$
through intermediate hidden states $\vec{h}_t \in \mathbb{R}^D$, using the
(discretized) state transition~\cref{eq:state_space_1,eq:state_space_2} with parameters
$\mat{A}, \mat{B}, \mat{C}, \mat{\Delta}$ and discretization
rule~\cref{eq:A_rule,eq:B_rule},
\begin{align}
    \vec{h}_{t}       & {}= \mat{\overline{A}} \vec{h}_{t-1} + \mat{\overline{B}} x_{t},\label{eq:state_space_1}                          \\
    y_{t}             & {}= \mat{C} \vec{h}_{t},\label{eq:state_space_2}                                                                  \\
    \widebar{\mat{A}} & {}= \exp(\mat{\Delta} \mat{A}),\label{eq:A_rule}                                                                  \\
    \widebar{\mat{B}} & {}= (\mat{\Delta} \mat{A})^{-1} \exp(\mat{\Delta} \mat{A} - \mat{I}) \cdot \mat{\Delta} \mat{B}.\label{eq:B_rule}
\end{align}

\subsection{SepMamba Architecture}\label{sec:networkarchitecture}

Our proposed SepMamba architecture operates on raw audio waveform, in contrast to
STFT-domain models such as TF-GridNet~\cite{tfgridnet} or SP-Mamba~\cite{spmamba}, and
is based on the U-Net~\cite{ronneberger2015unet} architecture --- composed of five
stages of down/up-sampling with a bidirectional Mamba (Bamba) block at each stage. Each
Bamba block additively combines the outputs of a stack of Mamba blocks with those of a
separate stack of Mamba blocks run on a reversed copy of the input,
\begin{equation}
    \mathrm{Bamba}(x) = \mathrm{Mamba}_1(x) + \mathrm{flip}(\mathrm{Mamba}_2(\mathrm{flip}(x))),
\end{equation}
where in each stage of processing we use the same number of Mamba blocks per Bamba
stack, but double the dimensionality of the Mamba layers (the channel dimension) with
every downsampling, and halve it with every upsampling. We use standard convolutions
for downsampling and matching transposed convolutions during upsampling. Skip
connections between features with different channel dimensions are projected to match
the target dimensionality using $1 \times 1$ convolutional layers. We use ReLU
activations throughout the whole network. For causal variants we match the number of
Mamba blocks per stage, but without reversing the inputs of either branch. The
architecture is illustrated in~\cref{fig:unet_structure}, and model configurations can
be found in~\cref{table:parameterization}.

Besides the small number of convolutions used for down/up-sampling, our model is the
first instance of a speaker separation network relying only on Mamba layers to learn
temporal dependencies, while other methods that have also started to incorporate Mamba
blocks --- such as SP-Mamba, which replaces the bidirectional LSTM module in TF-GridNet
with Mamba layers --- still rely on transformer-based layers for the bulk of their
computation.




\subsection{Experimental Setup}\label{sec:expsetting}

The models were trained and evaluated on the Wall Street Journal 0 (WSJ0) 2-speaker
setup~\cite{hershey2015deep} (WSJ0-2mix) using the dynamic mixing (DM) data
augmentation technique~\cite{zeghidour2020wavesplit}, which creates new mixtures from
randomly sampled speaker utterances on-the-fly. The sources are mixed by uniformly
sampling an SNR value from the interval $[-2.5, 2.5]$. Additionally, speed perturbation
is employed on the sources, changing the speed of the audio to between 95\% and 105\%
of the original. During training, gradient clipping of $5.0$ is used to ensure stable
convergence of the model parameters.

The model parameters are inferred by minimizing the negative Scale-Invariant
Signal-to-Distortion Ratio (SI-SDR)~\cite{si-sdr} and use utterance-level Permutation
Invariant Training~\cite{pit} (uPIT). During training, we threshold the loss at $-30$
as in WaveSplit~\cite{wavesplit}. The AdamW~\cite{adamw} optimizer is used with an
initial learning rate of $\num{15e-5}$, weight decay of $0.1$ and $\beta = (0.9,
0.999)$. Training loss was monitored for convergence at the initial constant learning
rate, upon which an exponential decay learning rate schedule is introduced with a gamma
value of $0.98$ to $0.99$, depending on the model. Training typically takes 6--7 days
on an NVIDIA A100 GPU.\@ A batch size of 1 was used for all training runs, as we found
larger batch sizes either produced significantly worse results or diverged entirely, a
phenomenon we also observed while attempting to replicate other models such as
SepFormer and SuDoRM-RF when trained using uPIT.

The SI-SDR improvement (SI-SDRi) is reported on the held-out test data, which measures
the difference between the SI-SDR after processing and the SI-SDR before processing. We
additionally report the Signal-Distortion Ratio improvement (SDRi) and Scale-Invariant
Signal-to-Noise Ratio improvement (SI-SNRi) to facilitate comparisons
in~\cref{table:result_comparison}.

We also report compute requirements in terms of giga-multiply-accumulates (GMAC) as a
hardware-independent measure of computational intensity, as well as wall-clock timings
specific to A100 GPUs and peak memory use.

\section{RESULTS AND DISCUSSION}

\renewcommand{\thefootnote}{\fnsymbol{footnote}}

\begin{table*}[t]
    \renewcommand{\arraystretch}{1.3}
    \begin{tabular}{ccccccccccc}
                   & \textbf{Size} & \textbf{Dim} & \textbf{\# Blocks} & \textbf{Kernel size} & \textbf{Stride} & \textbf{\# Params} & \textbf{Causal}\footnotemark[1] & \textbf{GMAC/s} & \textbf{SI-SNRi} \\
        \toprule
        {SepMamba} & S             & $64$         & $8$                & $16$                 & $2$             & $7.2$M             & \cmark                          & $12.46$         & $19.2$           \\
        {SepMamba} & S             & $64$         & $8$                & $16$                 & $2$             & $7.2$M             & \xmark                          & $12.46$         & $21.2$           \\
        {SepMamba} & M             & $128$        & $6$                & $16$                 & $2$             & $22$M              & \cmark                          & $37.0$          & $21.4$           \\
        {SepMamba} & M             & $128$        & $6$                & $16$                 & $2$             & $22$M              & \xmark                          & $37.0$          & $22.7$           \\

        \bottomrule
    \end{tabular}
    \centering
    \vspace{0.1cm}
    \caption{
        Parameterizations for the highlighted models. \textbf{Dim} refers to the
        dimensions of the Mamba blocks in the first and the last stage. \textbf{\#
        Blocks} refer to the total number of Mamba blocks per stage.
        }\label{table:parameterization}
\end{table*}

\begin{table*}[t]
    \renewcommand{\arraystretch}{1.3}
    \begin{tabular}{lccccccc}
                                                         & \textbf{SI-SNRi}       & \textbf{SI-SDRi} & \textbf{SDRi} & \textbf{\# Params} & \textbf{GMAC/s}     & \textbf{Fw. pass (ms)} & \textbf{Mem. Usage (GB)} \\
        \toprule
        {Conv-TasNet~\cite{convtasnet}}                  & $15.3$                 & $-$              & $15.6$        & $5.1$M             & $2.82$              & $30.79$                & $1.13$                   \\
        {DualPathRNN~\cite{luoDualpathRNNEfficient2020}} & $18.8$                 & $-$              & $19.0$        & $2.6$M             & $42.52$             & $101.83$               & $7.31$                   \\
        {SudoRM-RF~\cite{tzinisSudoRmRf2020}}            & $18.9$                 & $-$              & $-$           & $2.6$M             & $2.58$              & $69.23$                & $1.60$                   \\
        \midrule
        {SepFormer~\cite{sepformer}}                     & $20.4$                 & $-$              & $-$           & $26$M              & $257.94$            & $189.25$               & $35.30$                  \\
        {SepFormer~\cite{sepformer} + DM}                & $22.3$                 & $-$              & $-$           & $26$M              & $257.94$            & $189.25$               & $35.30$                  \\
        \midrule
        {MossFormer~\cite{mossformer} (S)}               & $-$                    & $20.9$           & $-$           & $10.8$M            & $-$\footnotemark[2] & $-$\footnotemark[2]    & $-$                      \\
        {MossFormer~\cite{mossformer} (M) + DM}          & $-$                    & $22.5$           & $-$           & $25.3$M            & $-$\footnotemark[2] & $-$\footnotemark[2]    & $-$                      \\
        {MossFormer~\cite{mossformer} (L) + DM}          & $-$                    & $22.8$           & $-$           & $42.1$M            & $70.4$              & $72.71$                & $9.57$                   \\
        {MossFormer2~\cite{mossformer2} + DM}            & $-$                    & $24.1$           & $-$           & $55.7$M            & $84.2$              & ${97.60}$              & $12.30$                  \\
        \midrule
        {TF-GridNet~\cite{tfgridnet} (S)}                & $-$                    & $20.6$           & $-$           & $8.2$M             & $19.2$              & $-$                    & $-$                      \\
        {TF-GridNet~\cite{tfgridnet} (M)}                & $-$                    & $22.2$           & $-$           & $8.4$M             & $36.2$              & $-$                    & $-$                      \\
        {TF-GridNet~\cite{tfgridnet} (L)}                & $-$                    & $23.4$           & $23.5$        & $14.4$M            & $231.1$             & $-$                    & $-$                      \\
        \midrule
        {SP-Mamba~\cite{spmamba}}                        & $22.5$\footnotemark[3] & $-$              & $-$           & $6.14$M            & $119.35$            & $148.11$               & $14.40$                  \\
        \midrule
        {\textbf{SepMamba (S) + DM} (ours)}              & $21.2$                 & $21.2$           & $21.4$        & $7.2$M             & $12.46$             & $17.84$                & $2.00$                   \\
        {\textbf{SepMamba (M) + DM} (ours)}              & $22.7$                 & $22.7$           & $22.9$        & $22$M              & $37.0$              & $27.25$                & $3.04$                   \\

        \bottomrule
    \end{tabular}
    \centering
    \vspace{0.1cm}
    \caption[hmm]{
        Performance comparison on the WSJ0-2mix dataset. GMAC/s is reported for a
        forward pass over 1 second of 8 kHz audio and calculated using
        ptflops\footnotemark[4]~\cite{ptflops}. Forward pass time is the average of 4 seconds of audio samples on 8 kHz. Memory
        usage is the peak memory usage during backpropagation of 4 seconds of audio
        sampled at 8 kHz. All calculations are in fp32.
    }\label{table:result_comparison}
\end{table*}

\Cref{table:result_comparison} compares the results achieved on the WSJ0-2mix dataset
by SepMamba and other prominent architectures. We highlight that SepMamba (M)
outperforms the transformer-based SepFormer and MossFormer (M) architectures with a
substantially reduced compute and memory footprint. In a similar manner, SepMamba (M)
also outperforms the related SP-Mamba model with significantly lower computational and
memory requirements, demonstrating the advantages of a fully Mamba-based architecture.
Lastly, our smaller model, SepMamba (S) outperforms prior models at a similar parameter
count.

Our models in the causal setting achieved SI-SNRi scores of $21.4$ and $19.2$. These
results outperform current state-of-the-art causal models, such as UX-NET (SI-SNRi
$13.6$)~\cite{uxnet} and Causal Deep Casa (SI-SNRi $15.2$)~\cite{causalcasa}.

\begin{figure}[t]
    \includegraphics[width=\linewidth]{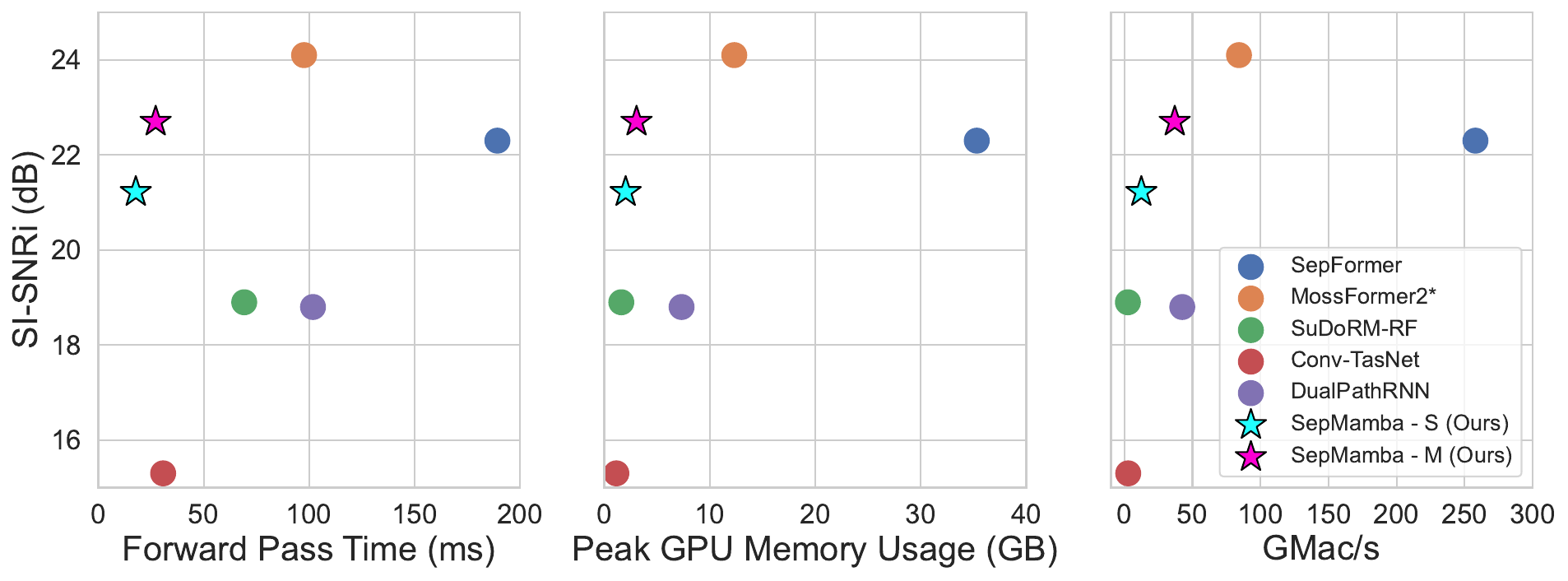}
    \caption{
        \textbf{(Left)} Average forward pass time on an NVIDIA A100 GPU for 4 seconds
        of audio samples at 8 kHz. \textbf{(Middle)} Peak GPU memory usage during the
        backpropagation of a 4 seconds sample at 8kHz on an NVIDIA A100 GPU.
        \textbf{(Right)} Multiply-accumulate (MAC) operations per seconds. *For
        MossFormer2 SI-SDRi is listed instead of SI-SNRi.\label{fig:main_comparison} }
\end{figure}

\noindent \textbf{Wall-clock timings}: \Cref{fig:main_comparison} (Left) shows the
average forward pass wall-clock time on an A100 GPU.\@ It is calculated by measuring
the number of forward passes each model can complete in 10 seconds and averaging. Both
SepMamba (S) and SepMamba (M) have significantly lower forward pass time than other
models while outperforming most of them in terms of SI-SDRi.

\noindent \textbf{Memory usage:} \Cref{fig:main_comparison} (Middle) shows the peak
memory usage by each model during backpropagation on a four second audio sample at $8$
kHz. Both SepMamba (S) and SepMamba (M) have memory use comparable to that of the
Conv-TasNet and SudoRM-RF, while being notably more efficient than MossFormer 2, and
significantly more efficient compared to SepFormer.

\noindent \textbf{Compute intensity:} \Cref{fig:main_comparison} (Right) shows the
number of giga-multiply-accumulates (GMAC) performed by each model during a forward
pass per second of audio sampled at 8kHz compared with SI-SNRi on WSJ0-2mix. SepMamba
(M) outperforms SepFormer with a fraction of the computational needs. While MossFormer2
achieves higher SI-SDRi, it requires more than twice the GMAC/s. Similarly, the smaller
SepMamba (S) has a strong performance with a relatively low computational overhead that
matches previous efficient models.

While in most cases SepMamba outperforms other methods at a lower compute and memory
footprint, several other methods achieve higher performance as a function of parameter
count. In the case where parameter count is the limiting factor on a system, we note
that there is plenty of opportunity for optimizing the parameter-efficiency of
SepMamba, such as parameter-sharing (e.g.\ between branches in Bamba stacks) or
adjusting the kernel size of the convolutions to trade-off compute and parameter
efficiency.

In our analysis we considered GMAC/s and wall-clock timings on A100 GPUs as the two
primary metrics for compute performance, but real-world performance depends heavily on
the characteristics of the system in question --- indeed the Mamba layer is a very
memory-bound operation with a low arithmetic intensity~\cite{mamba}, and so its runtime
on GPUs becomes dominated by transfers between local and global GPU memory, whereas the
attention mechanism still incurs quadratic compute cost, even when using the highly
optimized FlashAttention~\cite{flashattention} implementation. This implies that we may
expect an even greater performance advantage for Mamba-based architectures on systems
where less parallelism and compute is available.

A key advantage of Mamba is its compute and memory efficiency, especially over the
transformer-based attention mechanism. Another advantage is that it features a
substantially smaller state when operating in a recurrent, real-time context, as only
the current hidden state $\vec{h}_t$ is required to perform inference at the subsequent
timestep, compared to the attention mechanism which requires storing states
proportional to the entire length of the input sequence that it is attending to.

\subsection{Architectural Considerations}

In this section, we discuss how some of the architectural decisions influenced our
training runs. Decreasing the stride to two is beneficial, as the performance gained
outweighs the slightly longer forward and backward pass.

We also decided to construct our Bamba stacks with several blocks per branch in the
stack because pilot studies found that recombining the forward and reversed inputs
after every block leads to worse performance.

For the causal U-Net we additionally experimented with other activation functions as
well, namely SiLU~\cite{silu}, Mish~\cite{mish} and PRelu~\cite{prelu}, but we did not
see significant improvements in the pilot runs to justify further experimentation.


\footnotetext[1]{Refers to the causal setting of the Mamba blocks, not the
convolutions.}

\footnotetext[2]{Based on the descriptions in the public repository of
MossFormer~\cite{mossformer}, we were not able to obtain the source code for the
smaller models.}

\footnotetext[3]{Result taken from \url{https://github.com/JusperLee/SPMamba}.}

\footnotetext[4]{Mossformer GMAC/s calculation used FlopCounterMode from PyTorch with
the assumption $\text{MAC} = \text{FLOP}/2$. Mamba layer GMAC/s calculation is based on
\url{https://github.com/state-spaces/mamba/issues/110}.}

\section{CONCLUSION}



In this work we proposed SepMamba, a highly efficient U-net architecture for speech
separation based on Mamba layers, and demonstrated strong performance on
WSJ0-2mix using a fraction of the compute and memory budget of comparable methods.

We provided an extensive performance per compute review of recent methods, and find
that SepMamba broadly outperforms competing methods, especially at lower compute
budgets.

SepMamba forms a promising efficiency-focused alternative to transformer-based models,
suitable for use in lower-resource systems. Taken together with our strong causal
results, we suggest that SepMamba may be a strong candidate for real-world, low-power
and low-latency deployments.

\clearpage
\bibliography{refs}

\end{document}